%% file: main.tex
\documentclass[runningheads]{llncs}
\usepackage[T1]{fontenc}
\usepackage{graphicx}
\usepackage{amsmath}
\usepackage{float}
\usepackage{listings}
\usepackage{comment}
\usepackage{tikz}

\begin{document}

\title{OpenGL GPU-Based Rowhammer Attack (Work in Progress)}
%
%
\author{Antoine Plin\inst{1}\orcidID{0009-0005-4160-506X} \and
Frédéric Fauberteau\inst{2}\orcidID{0000-0002-1169-8040} \and
Nga Nguyen\inst{2}\orcidID{0000-0003-3273-8272}}
\authorrunning{A. Plin et al.}
%
\institute{Leonard de Vinci Graduate School of Engineering (ESILV), Paris, France\\
\email{antoine.plin@edu.devinci.fr} \and
De Vinci Higher Education, De Vinci Research Center, Paris, France
\email{\{frederic.fauberteau, nga.nguyen\}@devinci.fr}}
\maketitle              
\begin{abstract}
Rowhammer attacks have emerged as a significant threat to modern DRAM-based memory systems, leveraging frequent memory accesses to induce bit flips in adjacent memory cells. This work-in-progress paper presents an adaptive, many-sided Rowhammer attack utilizing GPU compute shaders to systematically achieve high-frequency memory access patterns. Our approach employs statistical distributions to optimize row targeting and avoid current mitigations. The methodology involves initializing memory with known patterns, iteratively hammering victim rows, monitoring for induced errors, and dynamically adjusting parameters to maximize success rates.

The proposed attack exploits the parallel processing capabilities of GPUs to accelerate hammering operations, thereby increasing the probability of successful bit flips within a constrained timeframe. By leveraging OpenGL compute shaders, our implementation achieves highly efficient row hammering with minimal software overhead. Experimental results on a Raspberry Pi 4 demonstrate that the GPU-based approach attains a high rate of bit flips compared to traditional CPU-based hammering, confirming its effectiveness in compromising DRAM integrity.
Our findings align with existing research on microarchitectural attacks in heterogeneous systems that highlight the susceptibility of GPUs to security vulnerabilities.
This study contributes to the understanding of GPU-assisted fault-injection attacks and underscores the need for improved mitigation strategies in future memory architectures.

\end{abstract}

\section{Introduction}

Modern computing systems increasingly rely on heterogeneous architectures, where CPUs, GPUs, and other specialized accelerators collaborate to deliver high-performance processing for numerous applications. While GPUs are widely known for their exceptional parallel computation capabilities, a growing body of research indicates that these accelerators can be repurposed for malicious tasks~\cite{Naghibijouybari2018,Naghibijouybari2022,Giner2024}.

A notable example is the \textbf{Rowhammer attack}, which exploits physical properties of Dynamic Random Access Memory (DRAM) cells through frequent row accesses.
DRAM organizes storage cells in a matrix of rows and columns, grouped into banks. Each DRAM cell consists of a capacitor and an access transistor,  with the capacitor storing a single bit of data as electrical charge.
Due to increasing density in modern DRAM chips, cells are physically closer together, making them more susceptible to electrical interference. When a row is accessed, the corresponding wordline is activated, causing a slight electrical disturbance in neighboring rows. With repeated activations (hammering), this disturbance can accumulate and discharge capacitors in nearby rows, flipping their stored values without directly accessing them~\cite{Kim2014}.

In CPU-based scenarios, repeated row activations exploit tight cell placement to induce charge leakage. However, \textbf{GPU-based} approaches can accelerate this significantly by leveraging massive parallelism and high memory bandwidth. This study demonstrates such a GPU-based Rowhammer attack on a Raspberry Pi 4.
For this, we use OpenGL compute shaders that allow general computations to be performed outside the traditional graphics rendering pipeline. 
We use statistical approaches and parallel GPU threads to intensify row hammering and increase the likelihood of bit flips in DRAM.

This paper makes several distinct contributions to the growing body of research on hardware security attacks:

\begin{itemize}
    \item \textbf{Novel GPU-based Rowhammer Implementation:} While Frigo et al.~\cite{Frigo2018} demonstrated microarchitectural attacks accelerated by integrated GPUs on mobile devices, our work specifically focuses on using OpenGL compute shaders for Rowhammer attacks on embedded systems like the Raspberry Pi 4, an approach not previously explored in the literature.
    
    \item \textbf{Advanced Statistical Hammering Patterns:} We extend the non-uniform access patterns introduced by Jattke et al.~\cite{Jattke2022} in their BLACKSMITH fuzzer, adapting these techniques to the GPU context with massively parallel thread execution for more effective DRAM disturbance.
    
    \item \textbf{Heterogeneous System Security Analysis:} Building upon the survey by Naghibijouybari et al.~\cite{Naghibijouybari2022}, we provide concrete evidence of cross-component vulnerabilities in heterogeneous systems, demonstrating how memory attacks can be launched from the GPU component against system-wide DRAM.
    
    \item \textbf{Practical Embedded Device Vulnerability:} While many Rowhammer studies focus on server or desktop systems, we demonstrate these attacks on the widely used Raspberry Pi platform, showing that even low-cost embedded devices with integrated GPUs are vulnerable to sophisticated memory corruption techniques.
    
    \item \textbf{Multi-sided GPU Hammering:} We adapt and improve upon the many-sided hammering technique from TRRespass~\cite{Frigo2020}, implementing it in the context of GPU compute shaders to bypass existing Target Row Refresh (TRR) mitigations in Low Power Double Data Rate 4 (LPDDR4) memory.
\end{itemize}

Our work bridges multiple research areas by combining GPU side-channel techniques~\cite{Naghibijouybari2018,Giner2024} with advanced Rowhammer methodologies~\cite{Frigo2020,Jattke2022} to create a more powerful attack vector that exploits the inherent parallelism of modern GPUs.

\section{Related Work}

Heterogeneous architectures combine CPUs, GPUs, and accelerators in a single platform for improved computational throughput. Despite performance gains, this integration broadens the attack surface by exposing shared memory resources and on-chip interconnects to various processes. Recent surveys~\cite{Naghibijouybari2022} highlight that attackers can exploit these shared resources for side-channel leakage, covert channels, or fault injection.

Rowhammer is a well-documented DRAM vulnerability \cite{Kim2014,Fraile2019,Mutlu2020} where repeated access to specific memory rows can induce bit flips in adjacent rows, leading to potential security breaches. Frigo et al.~\cite{Frigo2020} introduced ``TRRespas'', a method that exploits weaknesses in the TRR mitigation implemented in DDR4 DRAM modules. Their black-box Rowhammer fuzzer revealed that many DDR4 modules remain susceptible to Rowhammer attacks despite TRR defenses, emphasizing the need for more robust mitigation strategies.

TRR is a commonly implemented mitigation designed to counter Rowhammer attacks by monitoring memory access patterns and refreshing adjacent rows when a specific row is accessed excessively. However, TRR implementations are often proprietary and vary across manufacturers, leading to inconsistencies in effectiveness. Sophisticated attacks, such as TRRespass~\cite{Frigo2020}, have demonstrated the ability to bypass TRR by exploiting its blind spots, such as multi-sided hammering patterns.

There exists also other mitigation approaches such as Error-Correcting Code (ECC), Increased Refresh Rates and Row Randomization. ECC memory can detect and correct single-bit errors, reducing the impact of bit flips. However, ECC is not foolproof against multiple simultaneous bit flips. 
By refreshing DRAM rows more frequently with increased refresh rates, the likelihood of charge leakage causing bit flips is reduced. This approach, however, increases power consumption and may degrade performance. Randomizing the physical-to-logical row mapping makes it harder for attackers to target specific rows, though this requires hardware support. In future DRAM designs, emerging technologies, such as 3D-stacked DRAM or alternative memory architectures, aim to mitigate Rowhammer by reducing cell-to-cell interference. 

In software-based protection approaches, operating systems can implement memory allocation strategies to isolate sensitive data from vulnerable regions or detect abnormal memory access patterns.
Konoth {\it et al.} proposed ZebRAM that builds on virtualization extensions in commodity processors to isolate every DRAM row that contains data with guard rows to control data placement \cite{Konoth2018}.   
Despite all these mitigations, Rowhammer remains a persistent challenge, particularly as attackers develop more advanced techniques to exploit hardware vulnerabilities.

Concerning GPU side-channel attacks, 
Naghibijouybari et al.~\cite{Naghibijouybari2018} demonstrated the practicality of this kind of attacks by showing that an OpenGL-based spy can accurately fingerprint websites, monitor user activities, and infer keystroke timings. Additionally, they illustrated how a CUDA spy application could deduce internal parameters of neural network models used by other CUDA applications, highlighting vulnerabilities in shared GPU environments.

Similarly, Frigo et al.~\cite{Frigo2018} explored the security implications of integrated GPUs in mobile processors. They revealed that GPUs could accelerate microarchitectural attacks, such as side-channel and Rowhammer attacks, even from JavaScript within browsers. Their work emphasized the necessity for secure GPU design, especially as GPUs become more integrated into general-purpose computing.

Comparing to recent work based on CPUs such as AMD Zen-based platforms \cite{Jattke2024} or RISC-V \cite{Marazzi2024}, our GPU-based Rowhammer attack does not need to reverse engineer DRAM addressing function nor to synchronize with the refresh mechanism of the memory controller to bypass the TRR mitigations. Complex techniques such as scheduling flush and fence instructions to increase row activation throughput are replaced by using high-frequency memory access patterns with GPU compute shaders.  
We will show in the next section how we use the key factors such as high thread count,  memory bandwidth, compute shaders, bank-level parallelism, reduced CPU overhead, and randomized access patterns to 
amplify Rowhammer attacks on GPUs. 

\section{Implementation Details}

\subsection{High-Level Overview}

Our implementation aims to maximize DRAM activations from a GPU context. GPUs are inherently designed for massive parallelism, with thousands of threads executing simultaneously across multiple cores. This architecture provides a significant advantage for Rowhammer attacks compared to CPUs, which typically have fewer cores and threads. We rely on OpenGL ES 3.1
compute shaders to issue read operations on targeted rows. This modern GPU support  provides fine-grained control over memory access patterns that enables us to craft highly optimized hammering loops that maximize DRAM disturbance while avoiding detection by refresh mechanisms. The code is structured to:
\begin{enumerate}
    \item Allocate and initialize large GPU-accessible buffers or textures in DRAM;
    \item Issue ``hammer'' operations via compute shaders, which repeatedly read from chosen memory locations (victim rows);
    \item Vary row offsets randomly, creating multi-sided hammering patterns;
    \item Collect timing and error data via Shader Storage Buffer Objects (SSBOs) and host readbacks.
\end{enumerate}

Figure \ref{fig:logigram} illustrates the logical workflow of our GPU-based Rowhammer attack implementation, showing the main processing steps and decision points.

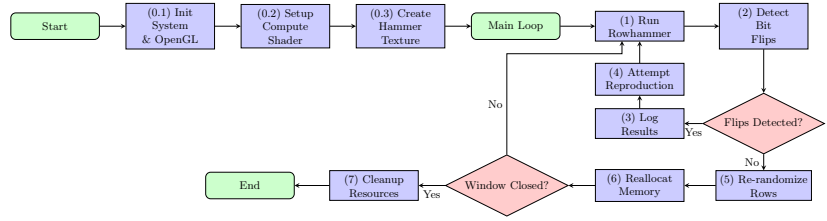
\begin{figure}[ht!]
    \centering
    \resizebox{\textwidth}{!}{%
    \input{figures/logigram.tex}
    }
    \caption{GPU-based Rowhammer Attack Logical Workflow} \label{fig:logigram}
    \label{fig:logigram}
\end{figure}

\subsection{Key Attack Primitives}

The following operations are important primitives for the attack.

\begin{itemize}
    \item \textbf{Parallel Threads:} We dispatch thousands of GPU threads (Fig.~\ref{fig:logigram}-(1)), each performing independent row accesses, thereby significantly increasing the total number of activations.
    \item \textbf{Adaptive Randomization:} Inspired by Blacksmith-style pattern fuzzing~\cite{Jattke2022}, we insert random offsets in row accesses (Fig.~\ref{fig:logigram}-(5)). This prevents simple row-refresh heuristics from detecting purely sequential patterns.
    \item \textbf{Repeated Reallocation:} To achieve broader DRAM coverage, we periodically reallocate textures (Fig.~\ref{fig:logigram}-(6)) so the GPU driver maps them to different physical memory regions.
    \item \textbf{SSBO Write-Back to Prevent Optimization:} Modern GPU compilers are highly optimized and may eliminate operations deemed unnecessary, such as repeated memory reads without observable side effects. To ensure the hammering loop is not optimized away, we accumulate the results of memory reads into a variable and write it back to a SSBO. This guarantees that the GPU executes the memory operations as intended, preserving the effectiveness of the attack.
\end{itemize}

To ensure that each row of the RGBA8 texture corresponds to a single row in DRAM, we carefully calculated the texture dimensions based on the DRAM row size and the format of the texture.
The Raspberry Pi 4 uses LPDDR4 memory, where each DRAM row typically has a size of 8KB (8192 bytes). An RGBA8 texture stores 4 bytes per pixel (1 byte for each channel: Red, Green, Blue, and Alpha). Therefore, the number of pixels per row in the texture must satisfy:

\[
\text{Pixels per row} = \frac{\text{DRAM row size (bytes)}}{\text{Bytes per pixel}} = \frac{8192}{4} = 2048
\]

Thus, each row of the texture should contain exactly 2048 pixels to align with a single DRAM row. The height of the texture can be chosen based on the total memory allocation required for the attack, ensuring that the texture fits within the available GPU-accessible DRAM. For example, a texture of size 2048x2048 would occupy:
\[
\text{Total size} = \text{Pixels per row} \times \text{Height} \times \text{Bytes per pixel} = 2048 \times 2048 \times 4 = 16 \text{ MB}
\]

This size is well within the 4GB LPDDR4 memory of the Raspberry Pi 4, leaving sufficient space for other system operations. By aligning the texture rows with DRAM rows, we ensure that memory accesses in the compute shader directly correspond to specific DRAM rows, maximizing the effectiveness of the Rowhammer attack.

\subsection{Compute Shader Implementation Analysis}

The compute shader implementation (Listing \ref{lst:gpu-code-snippet}) employs several key techniques to maximize the effectiveness of DRAM hammering.
{\scriptsize
\begin{lstlisting}[language=C++,
                 caption={Central GPU Compute Shader for Rowhammer Attack},
                 label={lst:gpu-code-snippet}
                ]
#version 310 es
layout(local_size_x = 16, local_size_y = 16) in;

// Uniforms controlling iteration count, victim row, etc.
uniform int uVictimRow, uBankStride, uMaxOffset, uIterations;
uniform uint uSeed;

// DRAM texture bound as readonly image
layout(rgba8, binding = 0) readonly uniform highp image2D uDRAMTexture;

// SSBO to store partial results (ensures read ops are done)
layout(std430, binding = 0) buffer OutputBuffer { vec4 outData[]; };

uint rand_lcg(inout uint seed) {
    seed = seed * 1664525u + 1013904223u;
    return seed;
}

void main() {
    ivec2 baseCoord = ivec2(gl_GlobalInvocationID.xy) * 4;

    // Per-thread seed for random offsets
    uint localSeed = uSeed
        + uint(gl_GlobalInvocationID.x) * 73856093u 
        + uint(gl_GlobalInvocationID.y) * 19349663u;

    // Bank-aware offset to stress DRAM interleaving
    int bankOffset = (baseCoord.x & 15) * uBankStride;
    int bankedX = bankOffset + (baseCoord.x & 511);
    vec4 accumulator = vec4(0.0);

    // Main hammering loop
    for (int i = 0; i < uIterations; i++) {
        int offset1 = int(rand_lcg(localSeed) % uint(uMaxOffset));
        int offset2 = int(rand_lcg(localSeed) % uint(uMaxOffset));

        int row1 = (uVictimRow + offset1) & (imageSize(uDRAMTexture).y - 1);
        int row2 = (uVictimRow - offset2) & (imageSize(uDRAMTexture).y - 1);

        ivec2 addr1 = ivec2((bankedX + i * 128) & 511, row1);
        ivec2 addr2 = ivec2((bankedX + i * 128 + 256) & 511, row2);

        // Repeated read operations
        vec4 val1 = imageLoad(uDRAMTexture, addr1);
        vec4 val2 = imageLoad(uDRAMTexture, addr2);

        // Accumulate the results to avoid compiler optimizations
        accumulator += (val1 + val2);
    }

    // Write final accumulator to SSBO
    uint index = gl_GlobalInvocationID.y 
               * (gl_NumWorkGroups.x * gl_WorkGroupSize.x)
               + gl_GlobalInvocationID.x;
    outData[index] = accumulator;
}
\end{lstlisting}
}
\subsubsection{Thread Organization}
The shader uses a 16×16 thread block configuration (\texttt{local\_size\_x = 16, local\_size\_y = 16}), creating 256 concurrent threads per work group. Multiple work groups can be dispatched in parallel, achieving thousands of simultaneous memory operations. Each thread hammers memory by reading specific DRAM rows in a loop, using random offsets.

\subsubsection{Random Offset Generation} The Linear Congruential Generator (LCG) function \texttt{rand\_lcg()} creates pseudo-random offsets for each thread:
    \begin{itemize}
        \item Each thread initializes a unique seed based on its global ID (a identifier unique to each thread) combined with the uniform \texttt{uSeed} value,
        \item Within the hammering loop, \texttt{offset1} and \texttt{offset2} are randomly generated for each iteration,
        \item These offsets modify the target row addresses, implementing the adaptive randomization pattern discussed earlier.
    \end{itemize}

\subsubsection{The Hammering Loop} The core of the attack is a tight loop that:
    \begin{itemize}
        \item Calculates two different row addresses (\texttt{row1} and \texttt{row2}) for each iteration,
        \item Uses bank-aware addressing (\texttt{bankedX}) to ensure hammering affects specific DRAM banks,
        \item Performs two image loads (memory reads) per iteration, targeting rows above and below the victim row,
        \item Accumulates the loaded values to prevent compiler optimization from eliminating the reads.
    \end{itemize}

\subsubsection{SSBO Result Collection} A SSBO stores the accumulated results:
    \begin{itemize}
        \item The \texttt{outData} buffer receives the final \texttt{accumulator} value from each thread,
        \item This serves two critical purposes: ensuring memory reads are not optimized away by the compiler, and allowing the CPU to verify shader execution,
        \item The (\texttt{index})  scheme ensures each thread writes to a unique buffer location.
    \end{itemize}
This implementation demonstrates how GPUs can be repurposed for memory attacks through their parallel architecture and direct DRAM access capabilities. The shader maximizes DRAM row activations by combining thread parallelism with randomized access patterns, all while avoiding driver or hardware optimizations that might otherwise reduce hammering effectiveness.

\section{Experimental Setup and Results}

\subsection{Hardware and Software Environment}
Our experiments target a Raspberry Pi 4 featuring a Broadcom BCM2711 SoC (Quad-core ARM Cortex-A72 CPU), a VideoCore VI GPU with support for OpenGL ES 3.1 compute shaders, and 4GB of LPDDR4 SDRAM.
We run a Raspbian-based OS, ensuring the necessary graphics drivers for GPU compute.

\subsection{Rowhammer Conditions and Reproduction Attempts}

By carefully tuning \(\texttt{uIterations}\) and launching multiple workgroups, we approach billions of DRAM accesses per second. The GPU’s parallel nature often surpasses what is achievable by CPU-based Rowhammer on the same board. LPDDR4 memory employs built-in refresh cycles and partial vendor-specific TRR logic. Our multi-sided, randomized approach is designed to elude simplistic refresh algorithms, consistent with findings from \cite{Jattke2022}.

To choose the victim rows, we randomize a “base” \(\texttt{uVictimRow}\) within the GPU memory space and let the compute shader offset it. This helps cover different physical rows, especially when combined with texture reallocation.
 
In addition, to confirm that flips are not spurious, the code implements reproduction attempts by: i) correcting the flipped bits in the texture; ii) re-running the same hammer parameters (\(\texttt{uIterations}, \texttt{uVictimRow}\), etc.); and iii) checking if the same bytes flip again, signifying a stable vulnerability in that DRAM region.
These steps ensure flips are truly due to Rowhammer rather than random DRAM errors or cosmic rays.

\subsection{Observed Bit Flips}

\begin{figure}
\includegraphics[width=0.8\textwidth]{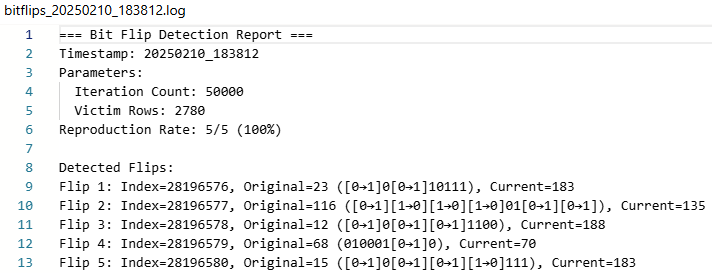}
\caption{An Extract of Log with Observed Bit Flips} 
\label{fig:log}
\end{figure}

We have carried out several runs lasting more than 20 hours each and without interruption in order to observe bit flips. This duration is necessary to bypass the ECC protection mechanism of the Raspberry Pi. The number of iterations is $5*10^4$ for these runs, and each thread performs 2 reads per loop iteration. With thousands of threads, total memory accesses can reach billions. The number of successful bit flips has been counted by comparing the original and modified (current) textures, as shown in Figure  \ref{fig:log}. 

We had on average 1024 bit flips for 5 different runs. More experiments should be done but from these runs we observe blocks of 32 rows with bit flips, and the addresses of these rows are continuous. The offset between the address of one block and the address of the next block is 16384 ($2^{14}$). This precise value might be related to the size of a subdivision of the memory or the DRAM addressing scheme of the GPU.     

It is noted that most flips occur in physically adjacent rows to the hammered addresses, consistent with classical Rowhammer theory. Some occasional flips, random outliers, appear in unexpected locations, potentially due to internal row remapping, temperature, or noise factors in DRAM.
Many flips can be reproduced across multiple runs, indicating stable vulnerabilities in specific DRAM regions.
An attacker could push the victim's data in these vulnerable locations using a method called Memory Massaging~\cite{Kogler2022} thus compromising its integrity.


\section{Discussion and Conclusion}


We have presented in this work-in-progress paper a GPU-based Rowhammer attack on the Raspberry Pi 4 using OpenGL ES compute shaders. By orchestrating large-scale, parallel DRAM accesses, our approach reliably induces bit flips in adjacent rows. We showed how randomization of row offsets, repeated reallocation, and a high level of parallelism can overcome naive refresh mechanisms, offering a new perspective on Rowhammer beyond CPU-centric methods. 

Our results confirm that GPU-based Rowhammer is not only feasible but can be highly effective under certain conditions. Even on a low-cost platform like the Raspberry Pi 4, the GPU can issue read operations at a rate sufficient to induce bit flips. This finding has broader implications: laptops, smartphones, or cloud servers—where GPUs or integrated graphics are widely used—may be susceptible to a similar approach if memory isolation and refresh strategies are insufficient. This attack may lead to other security implications and potential attack scenarios such as privilege escalation, denial-of-service or key manipulation. Indeed, bit flips could corrupt page table entries or other security-critical data structures, granting higher privileges to the attacker. Even if no direct privilege escalation occurs, repeated memory corruption could crash processes or the entire system, making it unavailable. In some setups, flipping bits in cryptographic keys stored in DRAM could weaken or fully compromise encryption. 


In order to continue our further research, several potential directions include:
\begin{itemize}
    \item \textbf{Adaptive Fuzzing Frameworks}: Automating row offset selection, iteration parameters, and temperature conditions to systematically discover the most vulnerable DRAM regions.
    \item \textbf{Cross-Architecture Validation}: Exploring identical GPU-based Rowhammer techniques on other SoCs, discrete GPUs, or platforms like x86 with integrated graphics.
    \item \textbf{Advanced Defences}: Developing real-time detection within the memory controller, using hardware performance counters or machine learning to flag suspicious GPU access patterns.
    \item \textbf{Scalable Attacks}: Testing whether multi-GPU setups in servers or HPC environments can amplify Rowhammer effects even further, especially if multiple accelerators share DRAM banks.
\end{itemize}

By highlighting the practical viability of GPU-driven Rowhammer, we emphasize the pressing need for robust hardware and software mitigations. These findings serve as a call to action for both the research community and hardware vendors to address Rowhammer in the context of heterogeneous and parallelized computing resources.




%
%
\bibliographystyle{splncs04}
\bibliography{main}

\end{document}

%% file: figures/logigram.tex
\usetikzlibrary{shapes.geometric, arrows, positioning}

\tikzstyle{startstop} = [rectangle, rounded corners, minimum width=2.5cm, minimum height=0.75cm, text centered, draw=black, fill=green!20]
\tikzstyle{process}   = [rectangle, minimum width=2.5cm, minimum height=0.75cm, text centered, draw=black, fill=blue!20]
\tikzstyle{decision}  = [diamond, aspect=2, text centered, draw=black, fill=red!20]
\tikzstyle{arrow}     = [thick,->,>=stealth]

\begin{tikzpicture}[node distance=1.25cm]

\node               (start)    [startstop] {Start};
\node[align=center] (init)     [process, right of=start, xshift=2cm]     {(0.1) Init\\System\\\& OpenGL};
\node[align=center] (shader)   [process, right of=init, xshift=2cm]      {(0.2) Setup\\Compute\\Shader};
\node[align=center] (texture)  [process, right of=shader, xshift=2cm]    {(0.3) Create\\Hammer\\Texture};
\node               (mainLoop) [startstop, right of=texture, xshift=2cm] {Main Loop};

\node[align=center] (hammer) [process,  right of=mainLoop, xshift=2.25cm] {(1) Run\\Rowhammer};
\node[align=center] (detect) [process,  right of=hammer,   xshift=2.25cm] {(2) Detect\\Bit\\Flips};
\node               (flips)  [decision, below of=detect,   yshift=-1.5cm] {Flips Detected?};

\node[align=center] (log)       [process, left of=flips, xshift=-2.25cm] {(3) Log\\Results};
\node[align=center] (reproduce) [process, above of=log]                  {(4) Attempt\\Reproduction};

\node[align=center] (rerandom) [process, below of=flips,   yshift=-0.5cm]  {(5) Re-randomize\\Rows};
\node[align=center] (realloc)  [process, left of=rerandom, xshift=-2.25cm] {(6) Reallocat\\Memory};

\node               (exit)    [decision,  left of=realloc, xshift=-2.5cm]  {Window Closed?};
\node[align=center] (cleanup) [process,   left of=exit,    xshift=-2.5cm]  {(7) Cleanup\\Resources};
\node               (end)     [startstop, left of=cleanup, xshift=-2.25cm] {End};

\draw [arrow] (start)    -- (init);
\draw [arrow] (init)     -- (shader);
\draw [arrow] (shader)   -- (texture);
\draw [arrow] (texture)  -- (mainLoop);
\draw [arrow] (mainLoop) -- (hammer);
\draw [arrow] (hammer)   -- (detect);
\draw [arrow] (detect)   -- (flips);

\draw [arrow] (flips.south) -- node[anchor=east]  {No}  (rerandom.north);
\draw [arrow] (flips.west)  -- node[anchor=north] {Yes} (log.east);
\draw [arrow] (log)         --                          (reproduce);
\draw [arrow] (reproduce)   --                          (hammer);
\draw [arrow] (rerandom)    --                          (realloc);

\draw [arrow] (realloc)    --                                                                (exit);
\draw [arrow] (exit.north) -- node[anchor=east]  {No} ++(0,2.85) -- ++(3.25,0) -- ++(0,0.36) (hammer);
\draw [arrow] (exit)       -- node[anchor=north] {Yes} (cleanup);
\draw [arrow] (cleanup)    --                                                                (end);

\end{tikzpicture}